\documentclass[11pt]{article}

\usepackage{amsmath}
\usepackage{amssymb}
\usepackage{latexsym}
\usepackage{graphicx}
\usepackage{psfrag,fancyhdr,epsfig}

\begin{document}

\begin{titlepage}

\begin{centering}

\hfill hep-th/yymmnnn\\

\vspace{1 in}
{\bf {SUPERSYMMETRIC INFLATION WITH THE ORDINARY HIGGS?} }\\
\vspace{1 cm}
{K. Tamvakis$^{1,2}$}\\
\vskip 0.5cm
{$^1 $\it{Physics Department, University of Ioannina\\
45110 Ioannina, GREECE}}\\
\vskip 0.5cm
{$^2$\it {Physics Department, CERN, CH-1211, Geneva 23,\\
Switzerland}}

\vspace{1.5cm}
{\bf Abstract}\\
\end{centering}
\vspace{.1in}

 We consider a model of $D$-term inflation in which the inflaton coincides with the standard Higgs doublet. Non-renormalizable terms are controlled by a discrete $R$-symmetry of the superpotential. We consider radiative corrections to the scalar potential and find that Higgs inflation in the slow-roll approximation is viable and consistent with CMB data, although with a rather large value of the non-renormalizable coupling involved.

\vfill

\vspace{2cm}
\begin{flushleft}

November 2009
\end{flushleft}
\hrule width 6.7cm \vskip.1mm{\small \small}
 \end{titlepage}

 Although inflation was first proposed within the framework of the gauge symmetry breaking phase transitions in the early universe\cite{GUTH}\cite{LIN}\cite{STEIN}, it was very soon realized that values of the mass and the self-coupling of the scalar field that drives inflation excluded the electroweak or other GUT Higgses and that inflation had to be associated with a separate sector only indirectly related with the rest of particle physics\cite{LINDE}. Independently of that, inflation has evolved into an exact science since its predictions for density fluctuations can be quantitatively tested against the very accurately measured CMB parameters. Recently, an interesting attempt was made to relate inflation to the electroweak Higgs in a version of the SM with a non-minimal Higgs coupling to the Ricci scalar\cite{SHAP}, nevertheless, no convincing way was found to avoid non-renormalizable terms that could destroy the required effective flatness of the scalar potential\cite{ESPINOZA}.

 In the present article we discuss whether it is possible that the Higgs doublet can be the driving field for inflation, namely the inflaton. We consider MSSM in the framework of $D$-term inflation. A pair of extra fields will be assumed to be present, neutral under the $SM$ gauge group but charged under the extra gauge symmetry, coupled to the Higgs doublets only through non-renormalizable terms. A discrete $R$-symmetry of the superpotential will also be assumed. The model will be studied in an expansion in the inverse Planck mass. We shall find that the model possesses the essential feature of $D$-term inflationary models, namely, a phase with the extra gauge symmetry unbroken and almost constant scalar potential. This is sufficient to initiate inflation. The additional feature of this model is that the inflaton in the final global vacuum phase is light possessing the flatness required by electroweak physics. We shall study slow-roll inflation in this model and find that Higgs inflation is viable and that it is achieved with inflaton field values below the Planck mass. Nevertheless, exact agreement with the value of the spectral index, requires either a rather large value of a particular coupling constant of a non-renormalizable term, or a less simplified version of the model.

Let us consider an extension of MSSM with an extra $U(1)$ gauge factor. All standard MSSM fields are assumed to be neutral under this new gauge group. We introduce only a pair of extra superfields $\phi_{\pm}$ charged with opposite charges $\pm 1$ under it.We also assume the presence of a non-zero Fayet-Iliopoulos parameter $\xi$ in the corresponding $D$-term $|\phi_+|^2-|\phi_-|^2+\xi$. This is the usual $D$-{\textit{term inflation}} set up\cite{D1}\cite{D2}. No renormalizable term between the MSSM fields and $\phi_{\pm}$ is possible due to $R$-parity, assuming that the new pair is even. Thus, the superpotential with all possible non-renormalizable terms that involve only the Higgs doublets are of the form
\begin{equation}
{\cal{W}}\,=\,\mu\,H\,H^c\,+\,\sum_{n,n'}\frac{\lambda_{n n'}}{M^{2n+2n'-3}}\left(HH^c\right)^n\left(\phi_+\phi_-\right)^{n'}\,.{\label{NRSUP}}
\end{equation}
Of course, we cannot solve the hierarchy problem and we shall just assume that $\mu$ is in the neighborhood of the electroweak scale. Terms like $(HH^c)^2$ or $\phi_+\phi_-$ can be forbidden with a suitable discrete $R$ symmetry. For example, we may assume that the superpotential wiil be invariant under the discrete $R$-symmetry\footnote{This is not a symmetry of the sector responsible for the breaking of supersymmetry and is broken when the latter is broken.} ${\cal{Z}}_3^{(R)}$
$$Q,\,L,\,N^c,\,U^c\,\rightarrow\,\alpha,\,\,\,\,\,\,{\cal{W}},\,H,\,D^c,\,E^c\rightarrow\,\alpha^2,\,\,\,\,\,\,\,\,H^c,\,\phi_+,\,\phi_-\rightarrow\,1\,.$$
This symmetry allows all standard renormalizable terms. The resulting superpotential is
\begin{equation}
{\cal{W}}\,=\,\mu\,H\,H^c\,+\,\frac{\lambda}{M}\left(HH^c\right)\left(\phi_+\phi_-\right)\,+\,\frac{\lambda'}{M^3}\left(HH^c\right)^3\,+\,
\frac{\lambda''}{M^3}\left(HH^c\right)\left(\phi_+\phi_-\right)\,+\,\dots.\,,
\end{equation}
where we do not show other fields\footnote{For example terms, like $\left(N^c\right)^2\phi_+\phi_-$, involving the right-handed neutrino superfield could play a role in inflation. This possibility is under exploration. Here, we shall assume that the values of parameters are such that these terms do not contribute to inflationary considerations.} apart from the Higgses and $\phi_{\pm}$. Throughout this article we shall assume that the defining scale of non-renormalizable terms will be of the order of the reduced Planck mass $M\,=\,M_P\,\sim\,2.4\times 10^{18}\,GeV$.

Since, we have considered non-renormalizable terms in the superpotential, we must do the same with the Kahler potential, which, up to $O(M^{-2})$, can be written as
$${\cal{K}}\,=\,{\cal{K}}_0\,+\,\frac{{\cal{K}}_1}{M^2}\,+\,O(M^{-4})\,.$$
The correction ${\cal{K}}_1$ contains a large number of terms but can be simplified without real loss of generality if we assume a discrete exchange symmetry of ${\cal{K}}$ between the two Higgs doublets. If that's the case, the direction $H\,=\,H^c$ satisfies electroweak $D$-flatness, since ${\cal{K}}_{H^c}H^c={\cal{K}}_HH$ along this direction.
Thus, assuming we stay along this direction, we may simplify the model even further replacing it with a model defined by a superpotential
\begin{equation}{\cal{W}}\,=\,\frac{\mu}{2}\,h^2\,+\,\frac{\lambda}{2M}h^2\phi_+\phi_-\,+\,\dots.\end{equation}
and a non-minimal Kahler term
\begin{equation}
{\cal{K}}_1\,=\,\,a|h|^4+b_+|\phi_+|^2|h|^2\,+\,b_-|h|^2|\phi_-|^2\,
+\,c_+|\phi_+|^4+c_-|\phi_-|^4\,+\,d|\phi_-\phi_+|^2\,+\,e|h|^2\phi_+\phi_-\,+\,h.c. \end{equation}
Next, we proceed to calculate the scalar potential to $O(M^{-2})$. We may also introduce a real canonical field $h\,=\,\phi/\sqrt{2}$, in terms of which, the scalar potential is
$$V\,\approx\,\frac{\mu^2}{2}\phi^2\,+\,\lambda\frac{\mu}{M}\phi^2\left(\phi_+\phi_-+\phi_+^*\phi_-^*\right)\,+\,\left(\frac{5}{16}-a\right)\frac{\mu^2}{M^2}\phi^4\,+\,\frac{\lambda^2}{M^2}\phi^2|\phi_+\phi_-|^2$$
$$\,+\,\frac{\lambda^2}{4M^2}\phi^4\left(|\phi_+|^2+|\phi_-|^2\right)\,+\,\frac{\mu^2}{2M^2}\left(1-b\right)\phi^2\left(|\phi_+|^2+|\phi_-|^2\right)\,+\,$$
$$-e\frac{\mu^2}{2M^2}\phi^2\left(\phi_+\phi_-+\phi_+^*\phi_-^*\right)\,+\,\frac{g^2}{2}\left(|\phi_+|^2-|\phi_-|^2\,+\,\xi\right)^2\,+\,$$
\begin{equation}
\frac{g^2}{M^2}\left(|\phi_+|^2-|\phi_-|^2+\xi\right)\left(2c_+|\phi_+|^4-2c_-|\phi_-|^4+\frac{b}{2}\phi^2\left(|\phi_+|^2-|\phi_-|^2\right)\right)\,.
\end{equation}
For simplicity, we have taken $b_+=b_-$. The $\phi$-dependent masses of $\phi_{\pm}$ are
\begin{equation}m_{\pm}^2\,=\,\frac{\lambda^2\phi^4}{4M^2}+\frac{\mu^2}{2M^2}(1-b)\phi^2\,\pm\sqrt{g^4\xi^2\left(1+\frac{b\phi^2}{2M^2}\right)^2+\frac{\mu^2\phi^4}{M^2}\left(\lambda-\frac{e\mu}{2M}\right)^2}\,.
\end{equation}
Note however that all $\mu$-dependent contributions are negligible. For example\footnote{The $U(1)$-breaking scale will be taken to be $O(10^{15}\,GeV)$.}, in the potential, the mass-term $\mu^2\phi^2$ is overwhelmed by $g^2\xi^2$ even for $\phi\sim O(M)$, if $\mu<<g\xi/M$. For $\sqrt{\xi}\,\sim\,10^{15}\,GeV$, this amounts to $\mu\,<<\,10^{12}\,GeV$, which is trivially satisfied. Similarly, in the expressions for $m_{\pm}^2$ the $\mu$-dependent terms $\delta m_{\pm}^2\,\approx\,\frac{\mu^2\phi^2}{2M^2}\left(1-b\pm\frac{\lambda^2\phi^2}{g^2\xi}\right)\,\approx\,\pm\frac{\lambda^2\mu^2\phi^4}{2M^2g^2\xi}$ lead to a negligible contribution $\delta m_{\pm}^2/m_{\pm}^2\,\approx\,\pm \mu^2/g^2\xi\,<<1$.

Neglecting $\mu$, the $\phi$-depended masses of $\phi_{\pm}$ are
\begin{equation}
m_{\pm}^2\,=\,\frac{\lambda^2\phi^4}{4M^2}\,\pm\,g^2\xi\left(1+\frac{b}{2}\frac{\phi^2}{M^2}\right)\,.
\end{equation}
Thus, both masses are positive, provided
\begin{equation}
\phi^2\,\geq\,\frac{2g}{\lambda}\sqrt{\xi}M\left(1+\frac{b}{2}\frac{\phi^2}{M^2}\right)^{1/2}\,
\approx\,\frac{2g}{\lambda}\sqrt{\xi} M\left(\,1\,+\,\frac{bg}{2\lambda}\frac{\sqrt{\xi}}{M}\right)\,\equiv\,\phi_c^2\,.\end{equation}
For $\phi\geq \phi_c$ the fields $\phi_{\pm}$ stay at the origin and the scalar potential is
\begin{equation} V(\phi)\,\approx\,\frac{g^2\xi^2}{2}\,+\,O(\mu^2)\phi^2\,,\end{equation}
which is quite flat. The global vacuum of the theory arises at $\phi<\phi_c$ and corresponds to
\begin{equation}\phi_+=0,\,\,\phi_-\approx\,\sqrt{\xi}\left(1\,-c_-\frac{\xi}{M^2}\right)\,.\end{equation}
The potential near the global minimum, apart from $O(\mu^2)$ terms, it aquires a term $\frac{\lambda^2\xi}{4M^2}\phi^4\,$, which is rather flat for $\lambda\sqrt{\xi}/M<<1$ as we will see.

It should also be noted that the standard soft supersymmetry breaking, introduced in the form quadratic mass-terms and the cubic scalar interactions appearing in the superpotential, is not going to have any effect on the flatness of the potential. For $m_s^2\phi^2$, the same argument applies as for the corresponding $\mu$-term mass, while for the $m_{\pm}$ masses, $m_s^2$ is negligible in comparison to $g^2\xi$. Finally, with the assumed superpotential, constrained by the given symmetries, no dangerous term arises. For instance, the largest such term is $\frac{m_s}{M3}h^6$ and it is irrelevant.

At the local minimum with $\phi_{\pm}=0$, the potential is independent of $\phi$ and radiative corrections become important. They are summarized in the Coleman-Weinberg formula, where only the contributions of $\phi_{\pm}$ appear, since these are the only fields that feel the effective supersymmetry breaking of the $D$-term. They are
$$\Delta V\,=\,\frac{1}{32\pi^2}\sum_{\pm}\left(\frac{m_{\pm}^4}{f_{\pm}^2}\ln\left(\frac{m_{\pm}^2}{f_{\pm}\Lambda}\right)\,-\frac{m_{\pm}^4(0)}{f_{\pm}^2}\ln\left(\frac{m_{\pm}^2(0)}{f_{\pm}\Lambda}\right)\,\right)\,,$$
where $m_{\pm}^2(0)$ are the masses of the fermions obtained by setting $\xi=0$. The function $f_{\pm}$ arises because of the non-minimal kinetic terms
$${\cal{K}}_i^{\,j}\left({\cal{D}}_{\mu}\Phi\right)^i\left({\cal{D}}^{\mu}\Phi\right)_j^{\dagger}\,\sim\,\left(1\,+\,\frac{b}{2}\frac{\phi^2}{M^2}\right)\left(|D\phi_+|^2\,+\,|D\phi_-|^2\right)\,+\,\dots.$$
Finally, we have
\begin{equation}
V(\phi)\,=\,\frac{g^2\xi^2}{2}\,+\,\frac{g^4\xi^2}{16\pi^2}\left(\,\ln\left(\frac{\phi^4}{\Lambda^4}\right)\,-\ln\left(1+b\frac{\phi^2}{2M^2}\right)\,\right)\,.{\label{RAD}}
\end{equation}
We have absorbed the constant part in a suitable rescaling of the renormalization scale.

Before we proceed further let's have a look at the different energy scales that appear in the problem. As a matter of fact we have already assumed, and made use of it, that the electroweak scale is negligible compared with the scale of the extra $U(1)$ breaking expressed by the parameter $\sqrt{\xi}$. For this scale there is the well-known cosmic string constraint\cite{COSMIC}, arising from the requirement that cosmic strings formed by $\phi_{-}$ at the end of inflation should be suppressed and will not affect the CMB data. This constraint reads
\begin{equation}
3.8\times 10^{15}\,GeV\,\leq \,\sqrt{\xi}\,\leq\,4.6\times 10^{15}\,GeV\,.
\end{equation}

 As long as we are in the phase with unbroken $U(1)$ the vacuum energy is approximately constant. This can lead to an inflationary phase in which the inflaton is identified with the Higgs field $\phi$. Let's proceed assuming the validity of the slow-roll approximation, namely
 $\ddot{\phi}<<H\,\dot{\phi}$ and $(\dot{\phi})^2<<V(\phi)$. The classical evolution equations in an FRW background are

 \begin{equation}
 3H\,\dot{\phi}\,\approx\,-\frac{V'(\phi)}{f(\phi)},\,\,\,\,\,\,H^2\,\approx\,\frac{V(\phi)}{3M^2}\,\Longrightarrow\,\frac{d\phi}{d\ln a}\,\approx\,-\frac{M^2}{f(\phi)}\frac{V'(\phi)}{V(\phi)}\,.
 \end{equation}
 The function
 $$f(\phi)\,=\,{\cal{K}}_h^h\,=\,1\,+\,2a\frac{\phi^2}{M^2}\,$$
 arises from the non-minimal kinetic terms. Integrating and taking the logarithm, we obtain
 \begin{equation}
 {\cal{N}}\,\equiv\,\ln\left(\frac{a_f}{a_i}\right)\,\approx\,-\frac{1}{M^2}\int_{\phi_i}^{\phi_f}d\phi\,f(\phi)\,\frac{V(\phi)}{V'(\phi)}\,\,,{\label{INT}}
 \end{equation}
 in terms of the {\textit{number of e-folds}} ${\cal{N}}$. Substituting the derivative of the potential
 $$V'(\phi)\,=\,\frac{4}{\phi}\left(\frac{g^4\xi^2}{16\pi^2}\right)\left(\frac{1+\frac{b}{4}\frac{\phi^2}{M^2}}{1+\frac{b}{2}\frac{\phi^2}{M^2}}\right)\,\approx\,\frac{4}{\phi}\left(\frac{g^4\xi^2}{16\pi^2}\right)\left(1\,-\frac{b}{4}\frac{\phi^2}{M^2}\right)\,,$$
 obtained from ({\ref{RAD}}) and integrating ({\ref{INT}}), we get
 \begin{equation}\frac{g^2}{\pi^2}{\cal{N}}\,\approx\,-\frac{4}{b}\left(1+\frac{8a}{b}\right)\ln\left(\frac{1-\frac{b}{4}\frac{\phi_i^2}{M^2}}{1-\frac{b}{4}\frac{\phi_f^2}{M^2}}\right)\,-\frac{8a}{b}\left(\,\frac{\phi_i^2}{M^2}\,-\frac{\phi_f^2}{M^2}\right)\,.{\label{CONSTR-1}}\end{equation}
 Taking ${\cal{N}}\,\approx\,60$ and making the generic coupling choice $g\,\sim\,0.1$ gives us $g^2{\cal{N}}/\pi^2\,\approx\,0.061$.

The comoving curvature perturbation is
\begin{equation}
{\cal{R}}_c\,=\,\frac{H^2}{2\pi|\dot{\phi}|}\,\approx\,\frac{V^{3/2}f}{2\pi\sqrt{3}M^3|V'|}\,\approx\,
\left(\frac{\pi}{g\sqrt{6}}\right)\left(\frac{\xi}{M^2}\right)\left(\frac{\phi_i}{M}\right)\left(\frac{1+2a\frac{\phi_i^2}{M^2}}{1-\frac{b}{4}\frac{\phi_i^2}{M^2}}\right)
\,.\end{equation}
 Matching this to the observed value ${\cal{R}}_c\,\approx\,4.7\times 10^{-5}\,$ amounts to the constraint
 \begin{equation}\left(\frac{\phi_i}{M}\right)\left(\frac{1+2a\frac{\phi_i^2}{M^2}}{1-\frac{b}{4}\frac{\phi_i^2}{M^2}}\right)\,\approx\,0.997\,.{\label{CONSTR-2}}
 \end{equation}
The two constraints ({\ref{CONSTR-1}}) and ({\ref{CONSTR-2}}) should allow us to obtain suitable $\phi_i$ and $\phi_f$ in terms of two "input" parameters $a$ and $b$. We shall limit ourselves to {\textit{subplanckian values}} of $\phi_i$. We may simplify (and restrict) the search by assuming that $a$ is subleading and taking it to be zero. In this case, solving ({\ref{CONSTR-2}}) we obtain
\begin{equation}
b\,\approx\,\frac{4M^2}{\phi_i^2}-\frac{4M}{\phi_i}\,.{\label{B}}
\end{equation}
 Varying $\phi_i/M\in [0.4,\,0.9]$, gives us values in the range $b\in [0.5,\,15]$. Note however that even $O(10)$ values give $b\phi^2/4<1$. Nevertheless, lowering the initial inflaton values increases the matching value of the parameter $b$.

The equation ({\ref{CONSTR-1}}) determines the value of $\phi_f$ at the end of inflation. As an example, for a characteristic pair of values we have
\begin{equation}\frac{\phi_i^2}{M^2}\,\approx\,0.25,\,\, \,\,\,\,\,b\,\approx\,8\,\Longrightarrow\,\frac{\phi_f^2}{M^2}\,\approx\,0.217\,.{\label{FIFAN}}\end{equation}
Assuming that inflation ends when the value $\phi_c$ is reached, we may identify $$\frac{\phi_f}{M}\,\approx\,\frac{\phi_c}{M}\,\approx\,\sqrt{\frac{2g}{\lambda}\frac{\sqrt{\xi}}{M}}\,.$$
 We see that the above choice ({\ref{FIFAN}}) corresponds to $\lambda\,\sim\,10^{-3}$.

Let us now consider the {\textit{slow-roll parameters}} $\epsilon,\,\eta$ and $\zeta$. They can be calculated in terms of the potential and its derivatives as
\begin{equation}\epsilon\,=\,M^2\left(\frac{V'}{V}\right)^2\,\approx\,\frac{1}{4}\left(\frac{M}{\phi}\right)^2\left(\frac{g^2}{\pi^2}\right)^2\left(1-\frac{b}{4}\frac{\phi^2}{M^2}\right)^2\,,\end{equation}
\begin{equation}\eta\,=\,2M^2\frac{V''}{V}\,\approx\,-\left(\frac{M}{\phi}\right)^2\left(\frac{g^2}{\pi^2}\right)\left(1+\frac{b}{4}\frac{\phi^2}{M^2}\right)\,
\end{equation}
\begin{equation}
\zeta^2\,=\,\frac{M^4}{4}\left(\frac{V'''V'}{V^2}\right)\,\approx\,\frac{1}{8}\left(\frac{M}{\phi}\right)^4\left(\frac{g^2}{\pi^2}\right)^2\left(1-\frac{b}{4}\frac{\phi^2}{M^2}\right)\,.
\end{equation}

Their values at the start of inflation are
$$\epsilon_i\,\approx\,\left(\frac{g^2}{2\pi^2}\right)^2,\,\,\,\,\,\,\,\,\,\,\eta_i\,\approx\,-\frac{g^2}{\pi^2}\left(\frac{2M^2}{\phi_i^2}-\frac{M}{\phi_i}\right)\,,$$
$$\zeta_i^2\,\approx\,\left(\frac{g^2}{\pi^2}\right)^2\left(\frac{M^3}{8\phi_i^3}\right)\,.$$
Note that at the end of inflation these parameters are still small. The {\textit{spectral index}} corresponding to these parameters is
$$n_s\,=\,1-6\epsilon_i\,+\,2\eta_i\,\approx\,1+2\eta_i\,=\,1-\,2\frac{g^2}{\pi^2}\left(\frac{2M^2}{\phi_i^2}-\frac{M}{\phi_i}\right)\,.$$
This is depicted in the figure below.

\centerline{\includegraphics[width=7cm]{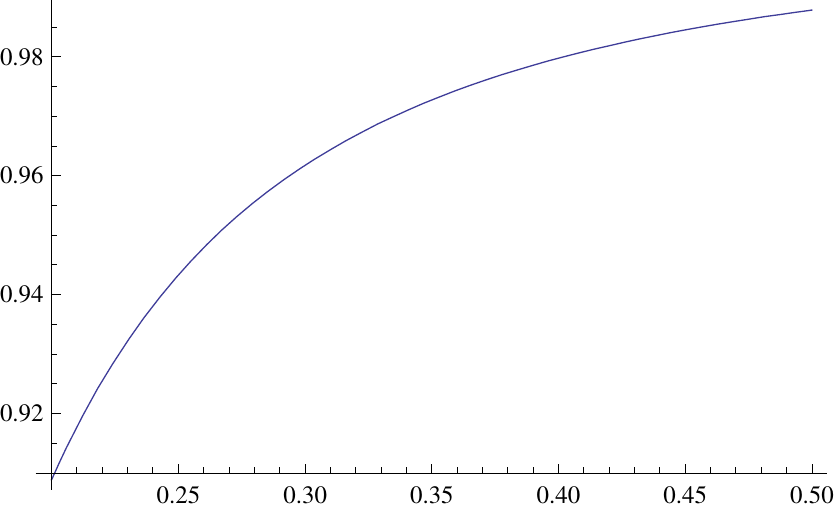}}

{\centerline{\textbf{ Plot of the spectral index in terms of $\phi_i/M$.}}}

Note that the smaller values of the spectral index are obtained for values of $\phi_i$ smaller than $0.5 M$ corresponding to $O(10)$ values of $b$. For example, $\phi_i\sim 0.3 M$ corresponds to $b\approx 15$. Note however that even such values have the appearing combination $b\phi_i^2/4<1$. Since the purpose of the present short article was to investigate whether Higgs inflation is viable, we shall not go here into a complete numerical study of the full parameter space.

 When the Higgs field $\phi$ has evolved to the critical value $\phi_c$ the theory makes a transition to the global minimum where the fields oscillate rapidly. The fields $\phi_{\pm}$ have a mass $g\sqrt{\xi}\,\sim\,4.6\times 10^{14}\,GeV$ and are directly coupled only to $\phi$. They can decay to MSSM matter, either gravitationally, or through their effective coupling $\lambda\sqrt{\xi}/M$ to $\phi$. The corresponding reheating temperature is
 $$T_R\,\approx\,\left(\frac{90}{\pi^2 g_{*}}\right)^{1/4}\left(\frac{\lambda^2 g}{8\pi^2}\right)^{1/2}\left(\frac{\sqrt{\xi}}{M}\right)^{3/2}M\,\approx\,10^{11}\,GeV\,.$$
 This is comparable to the reheating temperature corresponding to the gravitational decay rate to MSSM particles. As it stands the model requires an additional entropy dilution in order to circumvent entirely the gravitino problem.

Summarizing, let us briefly discuss the motivation and the main features of the model presented in the present article and, of course, the main result. The motivation is simply to investigate the possibility that the central scalar field of the Standard Model, namely the Higgs doublet, might be involved in inflation. The starting point had to be the MSSM because only supersymmetry could guarantee the required flatness of the inflaton potential. In order to avoid the $\eta$-problem of Supergravity, the framework of $D$-term inflation was considered and $MSSM$ was extended with an extra $U(1)$ gauge factor endowed with a non-zero Fayet-Iliopoulos term. Factors of this sort are not entirely uncommon in presently discussed effective particle models. Only a pair of extra fields were assumed to be present, neutral under the $SM$ gauge group but charged under the extra gauge symmetry. An important point is that the extra fields can couple to the Higgs doublets only through non-renormalizable terms. Finally, a discrete $R$-symmetry was also assumed for the superpotential. This model was studied in an expansion in the inverse Planck mass and was found to possess the essential feature of $D$-term inflationary models, namely, a phase with the extra gauge symmetry unbroken and almost constant scalar potential. The extra feature of this model is that the inflaton in the final global vacuum phase is light possessing the flatness required by electroweak physics. Slow-roll inflation was studied for the model and it was shown to occur for initial Higgs field values below the Planck scale.

\newpage
{\textbf{Acknowledgements}}

The author wishes to thank G. Dvali, M. Giovannini and A.Kehagias for discussions and the CERN Theory Group for hospitality. He also acknowledges the support of European Research and
Training Network MRTPN-CT-2006 035863-1 (UniverseNet).

\end{document}